# Matrix operators for complex interferometer analysis

Robert P. Dahlgren

UC Santa Cruz, Dept. of Electrical Engineering
1156 High Street, Santa Cruz, CA 95064-1077

Carl Sagan Center, The SETI Institute
189 Bernardo Ave., Mountain View, CA 94043-5203
(650) 810-0229   rdahlgren@seti.org

## ABSTRACT

A modeling methodology and matrix formalism is presented that permits analysis of arbitrarily complex interferometric waveguide systems, including polarization and backreflection effects. Considerable improvement results from separation of the dependencies on connection topology from the dependencies on the devices and their specifications. A non-commutative operator and embedding matrices are introduced allowing a compact depiction of the salient optical equations, and straightforward calculation of the amplitude and intensity transfer functions.

Keyword list: matrix multiplication, fiber optic sensors, gyroscopes, integrated optics, interferometers, optical design, metrology, tunable filters.

## 1. JONES MATRIX FORMALISM

Matrix methods are a powerful tool in optical engineering[1], and Jones matrices are particularly well-suited for modeling optical devices and systems having unidirectional quasi-plane wave propagation. Methodologies have been introduced leveraging matrix methods to simulate fiber optic resonators[2], micro-ring resonators[3], and other systems. Starting with the Jones matrix approach, consider a linear assembly of component matrices illustrated in Figure 1.

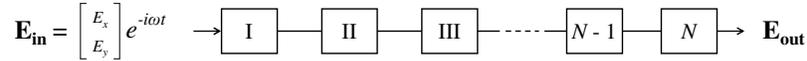

Figure 1. Optical component matrices representing an assembly of physical devices.

The optical train is represented by $N$ carefully-selected Jones matrices and rotation matrices[4] $\hat{J}_1$ through $\hat{J}_N$. These matrices usually represent a lumped-element portion of a component e.g. birefringent media, or optical effects occurring at a surface e.g. at a non-ideal fiber splice. In this sense number of "component" matrices might not be equal to the number of physical devices. For unidirectional propagation and two-port devices, the matrices are all 2×2 dimension, and operate on a 2-dimensional vector space having basis vectors (1,0) and (0,1) that represent the two orthogonal linear polarizations of the electric field. The amplitude transfer function of the system $\hat{J}$ is also a 2×2 matrix, being the ordered product of the component matrices

$$\hat{J} = \hat{J}_N \hat{J}_{N-1} \ldots \hat{J}_3 \hat{J}_2 \hat{J}_1 \qquad (1)$$

$\hat{J}$ can be thought of as a single "global" Jones matrix that describes the system. The matrix contents are generally weighted sums of exponential functions (or sines and cosines), whose complex arguments are functions of various phase factors, which themselves can be dependent on time, propagation distance, refractive index $n$, attenuation coefficient, and optical frequency $\omega_o$ to first order

$$\frac{\omega_o}{c}(n\,z) - i\omega_o t - \phi \qquad \text{where} \qquad \omega_o = \frac{2\pi c}{\lambda_o} \qquad (2)$$

where $\lambda_o$ is the wavelength of the quasi-monochromatic light, $\omega_o$ is the optical radian frequency, and $\phi$ is phase shift (if any) of a given component. The launched light into the first port in the optical train $\mathbf{E_{in}}$ is a 2×1 column vector representing the input E-field and its polarization state. The optical amplitude exiting the last port $\mathbf{E_{out}}$ is also a 2×1 column vector, equal to the system Jones matrix $\hat{J}$ multiplied by the input column vector

$$\mathbf{E_{out}} = \hat{J}\ \mathbf{E_{in}} \quad \text{and} \tag{3}$$

$$I_{out} = (\hat{J}_x\ \mathbf{E_{in}})(\hat{J}_x\ \mathbf{E_{in}})^* + (\hat{J}_y\ \mathbf{E_{in}})(\hat{J}_y\ \mathbf{E_{in}})^* \tag{4}$$

$$= (1,1)\ \{(\hat{J}\ \mathbf{E_{in}})(\hat{J}\ \mathbf{E_{in}})^*\} = (1,1)\ |\hat{J}\ \mathbf{E_{in}}|^2 \tag{5}$$

where output intensities are calculated for an ideal square law process. Equation (5) uses the (*1, 1*) row vector and dot product operation to sum the intensity of both polarizations in a compact manner. In time-independent or approximately time-independent situations, the $-i\omega_o t$ terms are commonly dropped from the calculations. If the responsivity $R(\omega)$ of the photodetector is known, then the ideal photocurrent $I_{PD}$ would be

$$I_{PD} = (1,1)\ \int R(\omega)\ |\hat{J}\ \mathbf{E_{in}}|^2\ \partial\omega \tag{6}$$

## 2. SCATTERING MATRIX FORMALISM

An interferometer subject to parasitic reflection, or an interferometer with a re-entrant topology such as an optical fiber gyroscope may be modeled with Jones matrices using topological "unfolding" techniques, but this approach is limited to less complex systems. Systems with bidirectional propagation may be faithfully simulated with scattering matrices[5,6], i.e. bidirectional Jones matrices, which are now generalized to include any arbitrary topology. This formalism starts with the identification and tabulation of the number and type of Jones matrices, the number *m* of unique ports, and port-to-port connections. Removing the two-port-per-device limitation means there may be 8×8 scattering matrices (representing a 4-port device such as a fiber-optic coupler) and 4×4 scattering matrices (for a 2-port device). The *N* scattering matrices are

$$\hat{S}_1,\ \hat{S}_2,\ \hat{S}_3,\ ...\ \hat{S}_{N-1},\ \text{and}\ \hat{S}_N \tag{7}$$

Unfortunately, this assemblage of matrices may not be chain-multiplied to obtain the system transfer function matrix. The second step of the formalism is the construction of a *global scattering* matrix $\hat{S}$, starting with a $2m \times 2m$ null matrix, and placing the *N* matrices along the diagonal as shown in Figure 2. The shaded regions represent the position of the scattering matrices corresponding to component I and component II within the larger matrix. If there is backreflection or other nonidealities in a given component, there would be additional nonzero coefficients in its submatrix.

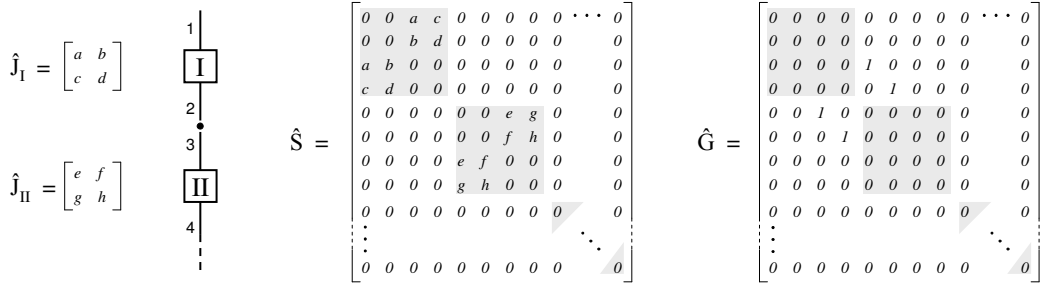

Figure 2. Construction of the global scattering matrix $\hat{S}$ from Jones matrices, and the matrix $\hat{G}$ connecting ports 2 and 3.

The resulting global scattering matrix $\hat{S}$ will relate all 2*m* output amplitudes to all input amplitudes, independent of the system interconnection topology, and will be block-diagonal and overdetermined. The vector space now has a dimension of 2*m*, and $\mathbf{E_{out}}$ and $\mathbf{E_{in}}$ are now $2m \times 1$ column vectors.

$$\mathbf{E_{out}} = \hat{S}\ \mathbf{E_{in}} \tag{8}$$

The third step is the construction of another $2m \times 2m$ matrix $\hat{G}$ according to the specific interconnection topology, again starting with a null matrix. The $\hat{G}$ matrix contains a quartet of unit coefficients at specific row-column combinations to represent each physical instance of interconnection from one component to another component.

$$\mathbf{E_{in}} = \hat{G}\ \mathbf{E_{out}} + \mathbf{E_o} \tag{9}$$

Figure 2 illustrates a portion of the $\hat{G}$ matrix showing an explicit connection between device I and device II, specifically between port #2 and port #3, and vise-versa. The final matrix $\hat{G}$ represents the entire system is a sparse matrix that is only dependent on topology and independent of ω and phase factors. $\mathbf{E_o}$ is another $2m \times 1$ column vector representing the light launched from an external source into the system. Both $\hat{S}$ and $\hat{G}$ have other attributes, such as symmetry to the extent they model reciprocal optical physics. Substituting equation (9) into equation (8) to eliminate $\mathbf{E_{in}}$, multiplying through by $\hat{S}^{-1}$, gathering terms, and solving for $\mathbf{E_{out}}$ results in

$$\mathbf{E_{out}} = \hat{S}\ \mathbf{E_{in}} = \hat{S}\ (\hat{G}\ \mathbf{E_{out}} + \mathbf{E_o}) \tag{10}$$

$$\hat{S}^{-1}\ \mathbf{E_{out}} = \hat{S}^{-1}\ \hat{S}\ (\hat{G}\ \mathbf{E_{out}} + \mathbf{E_o}) = \hat{G}\ \mathbf{E_{out}} + \mathbf{E_o} \tag{11}$$

$$\hat{S}^{-1}\ \mathbf{E_{out}} - \hat{G}\ \mathbf{E_{out}} = (\hat{S}^{-1} - \hat{G})\ \mathbf{E_{out}} = \mathbf{E_o} \tag{12}$$

$$\mathbf{E_{out}} = (\hat{S}^{-1} - \hat{G})^{-1}\ \mathbf{E_o} \tag{13}$$

Thus step four would be the computation of the amplitude transfer function matrix, called $\hat{H}$, that describes the $\mathbf{E_{out}}$ fields exiting the $2m$ ports as a function of the initial conditions $\mathbf{E_o}$

$$\hat{H} \equiv (\hat{S}^{-1} - \hat{G})^{-1} \tag{14}$$

$$\mathbf{E_{out}} = \hat{H}\ \mathbf{E_o} \tag{15}$$

It is important to note that this is a completely general approach and the degree of complexity or numbers of components is only limited by the ability to invert a large $2m \times 2m$ matrix. Equation (14) highlights the separation of topology and components, and permits valuable tradeoff studies to be performed during system architecture for applications such as fiber-optic sensors[7], integrated optic devices, and optical data storage[8].

### 3. OPERATOR MATRIX FORMALISM

Operators are widely used in quantum mechanics[9], optics[10], and in other disciplines; for example the $|\mathbf{E}|^2$ operator that is shorthand for $\mathbf{E}(\mathbf{E}^*)$. In the new formalism, the matrix $\hat{H}$ operates on the vector space containing the $\mathbf{E}$-vectors, and $\hat{S}$ and $\hat{G}$ does not commute in $\hat{H}$, except for trivial case of $\hat{S} = \hat{G} = \hat{I}$ the identity matrix. To eliminate unobservable or uninteresting variables, a rectangular matrix $\hat{A}$ having fewer rows than columns can be used to compact the global scattering matrix to a smaller scattering matrix.

$$\mathbf{E_{out}}' = \hat{A}\ \mathbf{E_{out}} \tag{16}$$

$$\mathbf{E_{out}}' = \hat{A}\ \hat{H}\ \mathbf{E_o} \tag{17}$$

$$\mathbf{E_{out}}' = \hat{A}\ \hat{H}\ \hat{A}^T\ \mathbf{E_o}' \tag{18}$$

$$\hat{H}' \equiv \hat{A}\ \hat{H}\ \hat{A}^T \tag{19}$$

where the prime indicates a "reduced" dimension and vector space, and the $^T$ indicates a transposed matrix. For example, it is often desirable to evaluate a system as a 2-port device, which necessitates reducing $\hat{H}$ to a 4×4 scattering matrix with a rectangular matrix $\hat{A}$ of dimension $4 \times 2m$. A reduced amplitude transfer function matrix $\hat{H}'$ may be used to compute the reduced-dimension output fields $\mathbf{E_{out}}'$ as in equation (15).

Alternately the system can be reduced even further to a unidirectional, 2×2 matrix, by using a different $\hat{A}$ matrix to extract the appropriate Jones matrix from $\hat{H}$, in this case the $\hat{A}$ matrix would be $2 \times 2m$. Taking input and output on the same physical port yields a set of $\hat{A}$ matrices

$$\hat{J}_{11}' = \hat{A}_1\ \hat{H}\ \hat{A}_1^T \qquad \text{where} \qquad \hat{A}_1 = \begin{bmatrix} 1 & 0 & 0 & 0 & \dots \\ 0 & 1 & 0 & 0 & \dots \end{bmatrix} \tag{20}$$

$$\hat{J}_{22}' = \hat{A}_2\ \hat{H}\ \hat{A}_2^T \qquad \text{where} \qquad \hat{A}_2 = \begin{bmatrix} 0 & 0 & 1 & 0 & \dots \\ 0 & 0 & 0 & 1 & \dots \end{bmatrix} \tag{21}$$

and so on. Given unidirectional optical propagation from the source into port $j$ toward the detector at port $k$, the system amplitude transfer function is described by the 2×2 matrix $\hat{J}_{kj}'$ given by[11]

$$\hat{J}_{kj}' = \hat{A}_k\ \hat{H}\ \hat{A}_j^T \tag{22}$$

For example, if the source is connected to reduced port #1 and the detector is connected to reduced port #2, then for this propagation direction the optical amplitude exiting port #2 is calculated by the reduced system Jones matrix $\hat{J}_{21}'$

$$\hat{J}_{21}' = \hat{A}_2\ \hat{H}\ \hat{A}_1^T \tag{23}$$

$$\mathbf{E_{out}}'_{(2)} = \hat{J}_{21}' \mathbf{E_o}'_{(1)} \tag{24}$$

This is a very compact yet complete expression of the optical system for the variables, states and ports of interest. The intensity is calculated as in equation (5), and is expressed most generally as

$$I_{out(k)} = (1, 1) \ |\hat{A}_k \ \hat{H} \ \hat{A}_j^T \ \mathbf{E_{in(j)}}|^2 \tag{25}$$

Consider the general case of the amplitude transfer function $\hat{H}(\omega)$ or reduced matrix $\hat{H}'(\omega)$ or reduced Jones matrix $\hat{J}_{kj}(\omega)$ for which exists the inverse Fourier transform (IFT), denoted by the symbol $\mathfrak{F}^{-1}$.

$$\hat{h}(\tau) = \mathfrak{F}^{-1}\{\hat{H}(\omega)\} = a \int \hat{H}(\omega) \ e^{i\omega\tau} \ \partial\omega \qquad \text{where} \qquad a \equiv (2\pi)^{-0.5} \tag{26}$$

The source has the amplitude spectrum described by the normalized function $F(\omega)$ for which also exists the IFT

$$f(\tau) = \mathfrak{F}^{-1}\{F(\omega)\} = a \int F(\omega) \ e^{i\omega\tau} \ \partial\omega \tag{27}$$

In the case of non-monochromatic light, the system response $\hat{H}_F$ would be the convolution of the scalar $F$ and the matrix $\hat{H}$, which is found by taking the Fourier transform $\mathfrak{F}$ of the product of the IFT of $F$ and the IFT of $\hat{H}$,

$$\hat{H}_F = F(\omega) \otimes \hat{H}(\omega) = \mathfrak{F}\{f(\tau) \ \hat{h}(\tau)\} = a \int f(\tau) \ \hat{h}(\tau) \ e^{-i\omega\tau} \ \partial\tau \tag{28}$$

$$I_{out(k)} = (1, 1) \ |a \int f(\tau) \ \hat{A}_k \ \hat{h}(\tau) \ \hat{A}_j^T \ e^{-i\omega\tau} \ \partial\tau \ \mathbf{E_{in(j)}}|^2 \tag{29}$$

Which for a monochromatic source $F(\omega) = \delta(\omega-\omega_o)$ is identical to equation (25). In conclusion, a methodology is presented which permits the computation of arbitrarily complex interferometers that includes all effects of polarization, reflection, source, and detector. This is expressed using a compact matrix operator formalism, which may computationally implemented on open-source or commercial numerical/symbolic software packages, and gain deeper insight into tradeoffs and architecture of interferometric sensors and systems.